\documentclass[%
 reprint,
superscriptaddress,
 amsmath,amssymb,
prl,
floatfix,
]{revtex4-2}

\usepackage{graphicx}
\usepackage{dcolumn}
\usepackage{bm}
\usepackage{amsmath,amssymb}
\usepackage{xcolor}
\usepackage{makecell}
\usepackage[detect-all]{siunitx}
\usepackage[normalem]{ulem}

\usepackage[colorlinks=false]{hyperref}
\usepackage{color,soul}
\usepackage{multirow}
\usepackage{enumitem}

\newcommand\figWidth{86}

\DeclareSIUnit\gauss{G}

\begin{document}

\preprint{PRL}

\title{Pair Correlations and Photoassociation Dynamics of Two Atoms in an Optical Tweezer}

\author{M.~Weyland}
\affiliation{The Dodd-Walls Centre for Photonic and Quantum Technologies, University of Otago, Dunedin, New Zealand}
\affiliation{Department of Physics, University of Otago, Dunedin, New Zealand}

\author{S.~S.~Szigeti}
\affiliation{Department of Quantum Science, Research School of Physics, The Australian National University, Canberra 2601, Australia}

\author{R.~A.~B.~Hobbs}
\affiliation{The Dodd-Walls Centre for Photonic and Quantum Technologies, University of Otago, Dunedin, New Zealand}
\affiliation{Department of Physics, University of Otago, Dunedin, New Zealand}

\author{P.~Ruksasakchai}
\affiliation{The Dodd-Walls Centre for Photonic and Quantum Technologies, University of Otago, Dunedin, New Zealand}
\affiliation{Department of Physics, University of Otago, Dunedin, New Zealand}

\author{L.~Sanchez}
\affiliation{The Dodd-Walls Centre for Photonic and Quantum Technologies, University of Otago, Dunedin, New Zealand}
\affiliation{Department of Physics, University of Otago, Dunedin, New Zealand}

\author{M.~F.~Andersen}
\email{mikkel.andersen@otago.ac.nz}
\affiliation{The Dodd-Walls Centre for Photonic and Quantum Technologies, University of Otago, Dunedin, New Zealand}
\affiliation{Department of Physics, University of Otago, Dunedin, New Zealand}

\date{\today}
\begin{abstract}
We investigate the photoassociation dynamics of exactly two laser-cooled $^{85}$Rb atoms in an optical tweezer and reveal fundamentally different behavior to photoassociation in many-atom ensembles. We observe non-exponential decay in our two-atom experiment that cannot be described by a single rate coefficient and find its origin in our system's pair correlation. This is in stark contrast to many-atom photoassociation dynamics, which are governed by 
decay with a single rate coefficient. We also investigate photoassociation in a three-atom system, thereby probing the transition from two-atom dynamics to many-atom dynamics. Our experiments reveal additional reaction dynamics that are only accessible through the control of single atoms and suggest photoassociation could measure pair correlations in few-atom systems. It further showcases our complete control over the quantum state of individual atoms and molecules, which provides information unobtainable from many-atom experiments.

\end{abstract}

\maketitle

Chemical processes govern the natural world and are used to create desired molecular structures. Such reactions usually occur in macroscopic samples of atoms and molecules that interact in many different ways. However, the tantalizing prospect of assembling individual molecules atom-by-atom via optical tweezers is emerging~\cite{Ashkin_Optical_Tweezer, Harvard_single_atom_PA, Liu_2019, Zhang_2020}. Developed to its full capacity, this bottom-up approach could realize the enduring scientific ambition of arranging atoms in molecules the way we want~\cite{Feynman_talk, Krems_2008, Ospelkaus_2010}. Furthermore, studying the formation of individual molecules isolates the reaction dynamics of interest from additional undesirable processes, such as spurious inter-molecular collisions, thereby giving unprecedented insight into the underlying physics.

An ideal process for controlled molecular formation is photoassociation, where light converts two colliding atoms into a molecule. Recent experiments have shown the formation of a single molecule via photoassociation~\cite{Harvard_single_atom_PA, Liu_2019} and magnetoassociation~\cite{Zhang_2020} of exactly two atoms. However, there has only been one prior study into the dynamics when photoassociating exactly two atoms~\cite{Sompet2013}, and its use of near-resonant light resulted in strong photon scattering, which made the atomic dynamics between collisions effectively classical.

Quantum correlations can change the photoassociation rate in many-atom systems~\cite{Kinoshita_2005}. Moreover, theoretical studies show that the photoassociation process itself can affect atom-atom correlations, leading to complex dynamics~\cite{Koch_2009, Gasenzer_2004, Holland_2001, Naidon_2006}. However, to date all experimental studies of photoassociation dynamics are well-described by a single time-independent rate coefficient, indicating that photoassociation did not affect atom-atom correlations~\cite{Schloeder_2002, McKenzie_2002, Prodan_2003, Dutta_2014, Wester_2004, Jones_PA_review}.

In this work, we observe the quantum dynamics of exactly two atoms undergoing photoassociation in an optical tweezer. The dynamics differs profoundly from those observed in many-atom ensembles. In particular, the dynamics is more complex and molecule formation cannot be described by a single rate coefficient. This is due to the pair correlation in the two-atom system; for two atoms the center-of-mass and relative-position degrees of freedom are separable in an optical tweezer. Consequently, thermally-populated relative-position states either possess strongly positive pair correlations (with a high chance of finding the two atoms close together) or a node in the pair correlation function at zero interatomic separation (i.e. the atoms are anti-correlated and there is a low chance of finding the pair close together).
Anti-correlated states are unaffected by photoassociation on short timescales, whereas relative-position eigenstates with strongly positive pair correlations lead to fast molecule formation.
We confirm that this is the underlying cause by investigating the photoassociation dynamics of three atoms, which approach the well-known dynamics of many-atom experiments.
\begin{figure}
  \includegraphics[width=\figWidth mm]{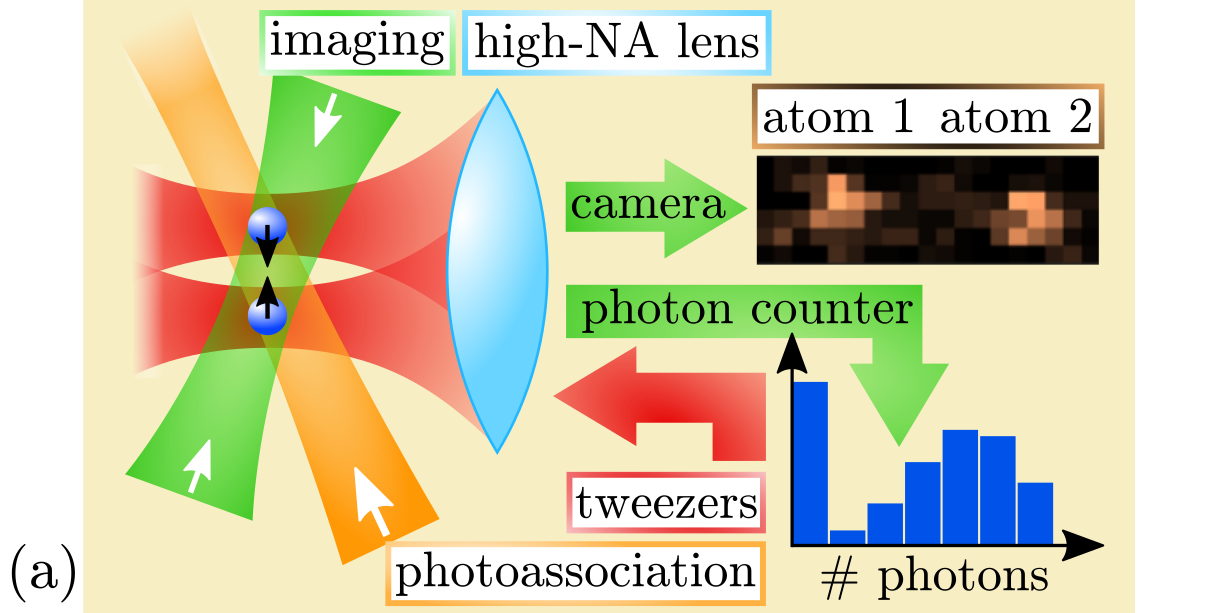}
  \newline
    \includegraphics[width=\figWidth mm]{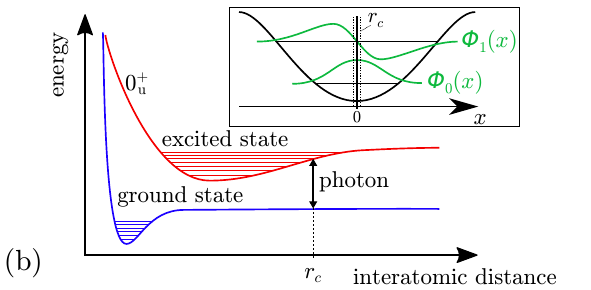}
	\caption{(a) Schematic of experimental setup and measurement. Tweezer light is focused by a high-NA lens (NA = 0.55) to a radial diameter of \SI{1.1}{\micro \metre}. Atoms in the tweezers are imaged using an EMCCD camera and merged into a single optical tweezer (black arrows). After applying photoassociation light, a single photon counter measures the population of the tweezer. Green and orange beams show imaging and photoassociation light, respectively. 
	(b) Sketch of the photoassociation process. Inset: 1D slice of the two lowest energy relative-position eigenstates of two atoms in our optical tweezer, with Condon radius $r_c$ marked for comparison.
	\label{fig:exp_visualization}}
\end{figure}

\emph{Experiment.---} We prepare a single $^{85}$Rb atom in an optical tweezer with an efficiency of around 80\% and detect it using an EMCCD camera~\cite{Gruenzweig2010}. The single atoms are optically pumped into \mbox{$|$F\,=\,2,\,m\,=\,-2$\rangle$} in a \SI{8.7}{\gauss} magnetic quantization field. By adiabatically merging two spatially-separated tweezers that each contain a single atom, we obtain one tweezer with exactly two atoms at a peak density of \SI{1.7e13}{\per\cubic\centi\meter}~\cite{Sompet_2019}.
Figure~\ref{fig:exp_visualization}(a) shows the central components of the setup. We use a high-numerical aperture lens (NA = 0.55) to create the tightly-focused light of the optical tweezers (red beams), as well as to collect a large proportion of the scattered imaging light (green beam), which is sent to the camera to confirm atom capture in both tweezers.

After transferring both atoms to the same tweezer, the atoms are exposed to photoassociation light at a frequency near \SI{377.00114}{\tera\hertz}, \SI{106}{\giga\hertz} red-detuned from the atomic D1 transition. Using a stable cavity we can reproduce this frequency with a precision of \SI{2}{\mega\hertz}.
A Ti:Sapphire laser delivers this light in \SI{140}{\nano\second}-duration pulses, during which the optical tweezer is turned off to eliminate any light shifts from the tweezer. 
Photoassociation dynamics occur on timescales from several \SI{}{\micro\second} to several \SI{}{\milli\second}. We form molecules in a high vibrational level of a $0^+_u$ state [Fig.~\ref{fig:exp_visualization}(b)], with the target state determined by the polarization and frequency of the photoassociation light~\cite{Heinzen_PA_Rb_1993, Heinzen_PA_Rb_1994, Bergeman_2006, Jones_PA_review, Degenhardt_2003}.

We detect a photoassociation event by imaging the tweezer after a given time of exposure to photoassociation light. Formed molecules either quickly decay to the groundstate, which does not scatter imaging light, or back into two atoms, which have now received enough energy to escape the trap. In both cases no atoms remain in the tweezer. We use a single photon counter to precisely measure the amount of scattered imaging light, which allows us to determine the number of atoms in a single tweezer~\cite{Reynolds2020}.

\emph{Pair correlation for two atoms.---} Since the optical tweezer is well-approximated as a harmonic potential, the two-atom centre-of-mass and relative-motional degrees of freedom are separable. Our experiments are performed with identical bosonic $^{85}$Rb atoms, so the even-parity eigenstates
\begin{align}
\phi_\textbf{n}(\textbf{r}) = \varphi_{n_x}(x)\varphi_{n_y}(y)\varphi_{n_z}(z)
\label{eq:3DEigenstates}
\end{align}
of the Hamiltonian
\begin{align}
H_\text{rel}(\textbf{r}) = -\frac{\hbar^2}{2\mu}\nabla_\textbf{r}^2 + \sum_{i=x,y,z} \frac{1}{2} \mu \omega_i^2 r_i^2,
\label{eq:Hamiltonian}
\end{align}
form a complete set for the dynamics. Here, $\textbf{r} = (x,y,z)$ is the relative-position coordinate and $\varphi_{n_i}(r_i)$ are eigenstates of a 1D harmonic oscillator with frequency $\omega_i$ and mass $\mu = m_\text{Rb}/2$, with $m_\text{Rb}$ the mass of a rubidium atom (Fig.~\ref{fig:exp_visualization}(b) inset). To ensure that $\phi_\textbf{n}(\textbf{r})$ is symmetric under particle exchange, $(-1)^{n_x+n_y+n_z}=1$. At zero separation between the atoms, these eigenstates either have a peak ($n_x,n_y,n_z$ all even) or a node (two of $n_x,n_y,n_z$ odd, one even). Atom pairs thermally populate these relative-position eigenstates.
Although including a finite-range atom-atom interaction modifies the eigenstates, the reflection symmetries of $H_\text{rel}(\textbf{r})$ persist for a spherically-symmetric interaction.
Therefore, the classification of two-atom states into strongly positive pair correlated (peaked) and anti-correlated (nodal) remains valid even in the presence of realistic interactions.

The pair-correlations of the two-atom wavefunction strongly determine the photoassociation dynamics~\cite{Koch_2009, Koch_2012}. The strength of the photoassociation process depends on the wavefunction at the Condon radius $r_c$, the interatomic distance at which photoassociation light can resonantly transfer atoms to an electronically-excited molecular potential~\cite{Heinzen_PA_Rb_1993}. In our experiment, $r_c=$ \SI{4.3}{\nano\metre}~\cite{Bergeman_2006}. For laser-cooled atoms, the thermal de Broglie wavelength $\lambda_\text{dB}$ is large compared to $r_c$, so the wavefunction at the Condon radius is either close to zero for anti-correlated nodal states or close to a maximum for strongly positive pair-correlated peaked states (Fig.~\ref{fig:exp_visualization}(b) inset). Consequently, atom pairs initially prepared in peaked eigenstates exhibit a much faster photoassociation rate compared to atom pairs initially prepared in nodal states.
Therefore, photoassociation in the two-atom system requires at least a \emph{two-timescale model} that accounts for the formation of molecules at two different rates. Contrast this to the many-atom case, which is described by a one-timescale model~\cite{McKenzie_2002}.

\begin{figure}
\includegraphics[width=\figWidth mm]{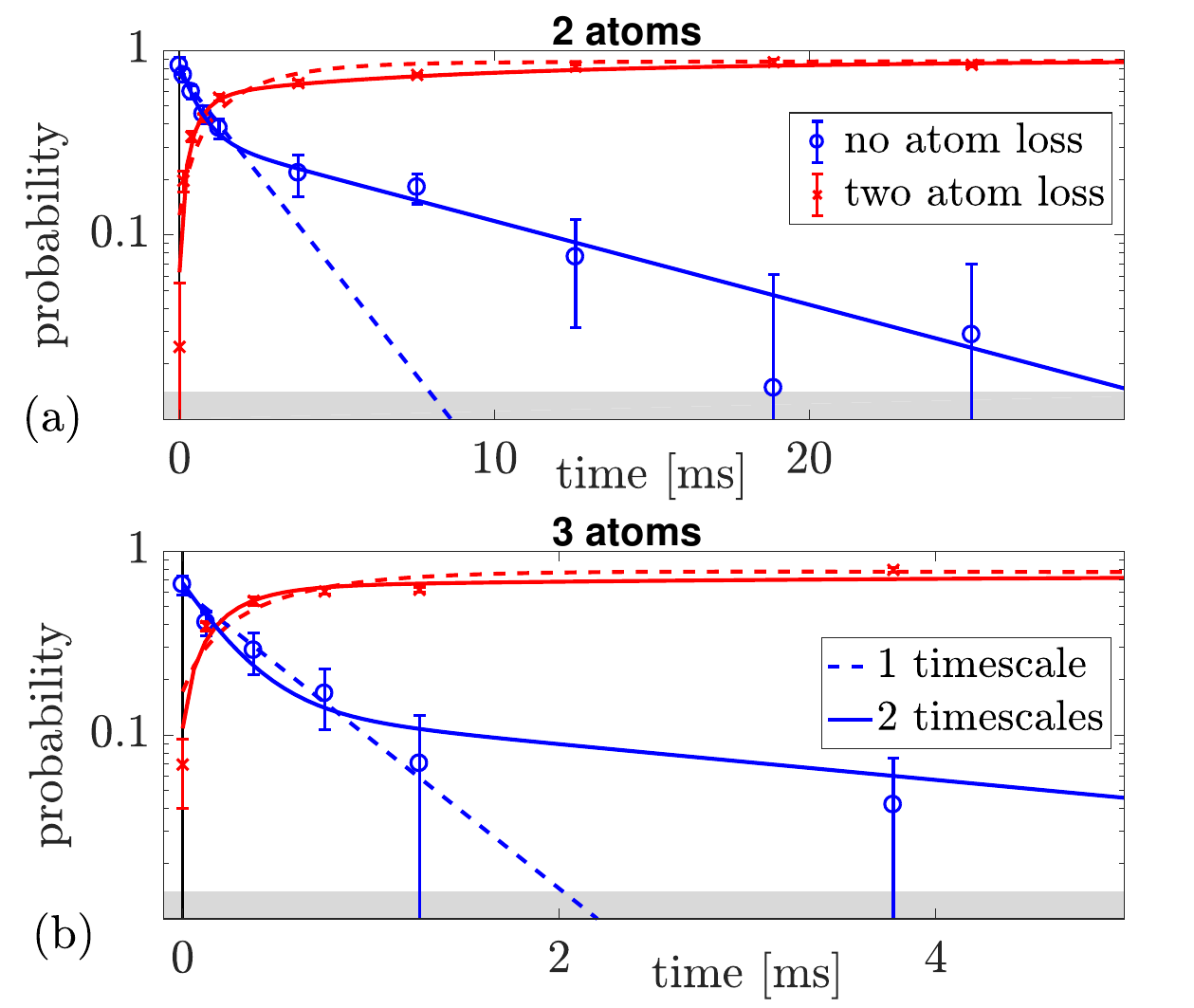}
	\caption{Evolution of the tweezer's population as a function of photoassociation time for two atoms (a) and three atoms (b) at \SI{35}{\micro\kelvin}. Given the measured radial and axial trapping frequencies of 88~kHz and 14~kHz, respectively, this temperature corresponds to mean relative-motional mode occupations of $\bar{n}_r \sim 7$ in the radial directions and $\bar{n}_z \sim 40$ in the axial direction. In both panels, dashed and solid lines indicate the best one- and two-timescale fits to the experimental data, respectively. The grey area marks our detection sensitivity limit.
	Since the three-atom system has three atom pairs, compared to one in the two-atom system, we expect a decay timescale three-times faster for the same rate coefficient.}
	\label{fig:PA_all_evolution}
\end{figure}

\emph{Two-timescale model of photoassociation.---} Figure~\ref{fig:PA_all_evolution}(a) shows a typical measurement of the probability of finding zero or two atoms in the optical tweezer as a function of photoassociation time. Red crosses show the probability that both atoms are lost due to photoassociation, while blue circles indicate the probability that both atoms remain. The one-timescale fit (dashed line) shows that the photoassociation dynamics cannot be reproduced using only one photoassociation rate. However, we obtain a good reproduction of the observed dynamics with a two-timescale model (solid line) that fits two independent two-atom populations with separate photoassociation rates~\cite{Suppl}. 

Using a $\chi^2$~test~\cite{Hughes_Hase_Uncertainties} on all rate measurements, we find that the one-timescale model is clearly rejected, while the two-timescale model is accepted in most cases~\cite{Suppl}. Our statistical analysis shows the importance of atom-pair correlations on the photoassociation dynamics.

Further evidence for our two-timescale model is provided via a numerical simulation of the relative coordinate wavefunction $\psi(\textbf{r},t)$. Since photoassociation principally occurs near the Condon radius $r_c$~\cite{Heinzen_PA_Rb_1993}, we use a simplified model that treats photoassociation as an absorbing hard-shell potential of strength $\hbar \Gamma$ and width $w$, located at relative distance $r_c$~\cite{Suppl}:
\begin{equation}
    i \hbar \frac{\partial}{\partial t} \psi(\textbf{r},t) = \left[H_\text{rel}(\textbf{r}) - i V_\text{PA}(\textbf{r}) \right] \psi(\textbf{r},t), \label{Eq:SE}
\end{equation}
where 
\begin{equation} \label{Eq:hard_sphere}
    V_\text{PA}(\textbf{r}) = \begin{cases} 
                                \hbar \Gamma, & r_c - w \leq |\textbf{r}| \leq r_c, \\
                                0,   & \textrm{otherwise}.
   \end{cases}
\end{equation}
We use $w = 0.38$\,pm, corresponding to the interatomic distances accessible at the observed 28\,MHz resonance width. Simulations are conducted by averaging over a thermal ensemble of initial states evolved under Eq.~(\ref{Eq:SE}). The initial states are relative-position eigenstates $\phi_\text{n}(\textbf{r})$ of Eq.~(\ref{eq:Hamiltonian}).

\begin{figure}
\includegraphics[width=\figWidth mm]{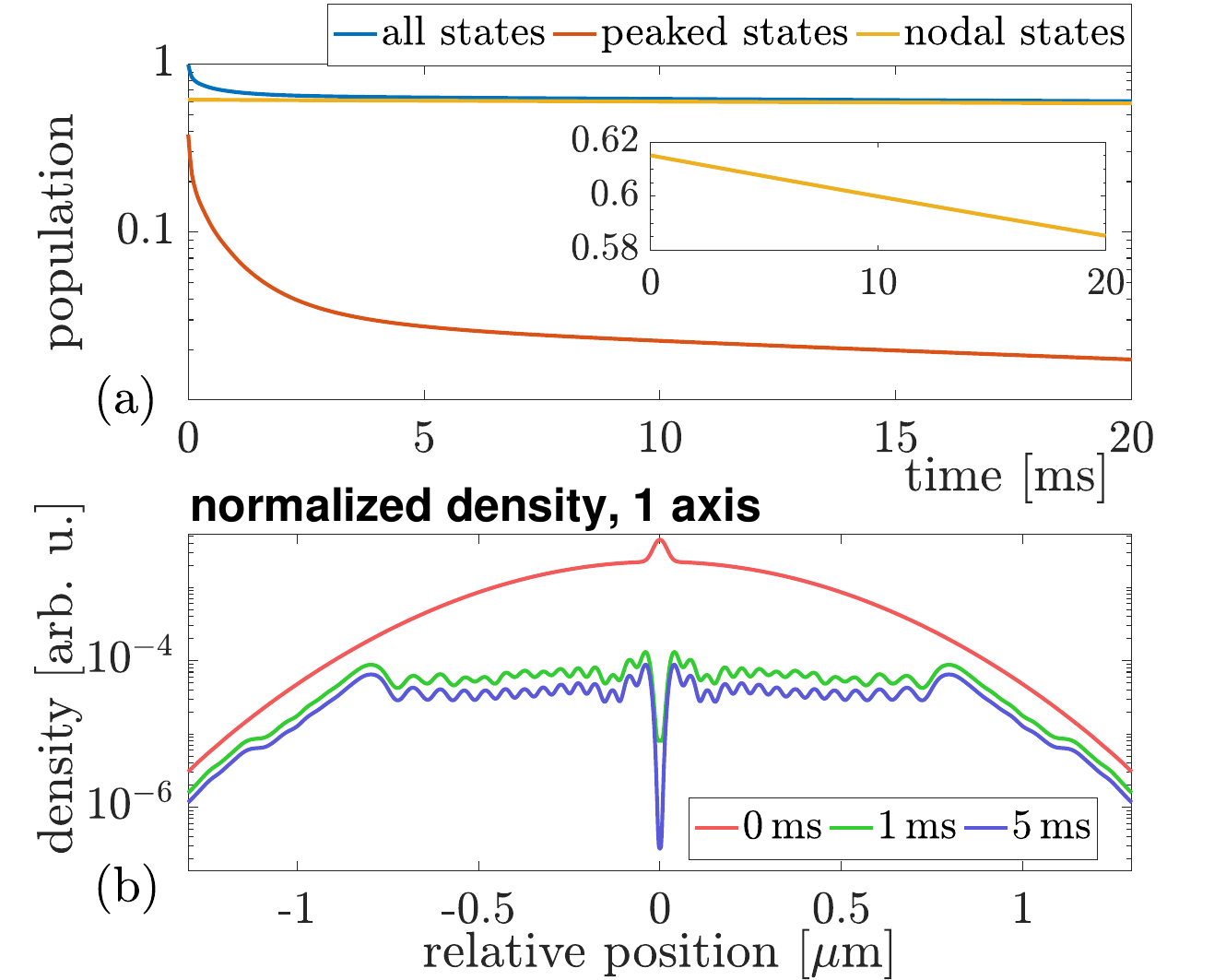}
	\caption{Numerical simulation of the atom-pair evolution in the tweezer under photoassociation. (a) Population over time for an ensemble at \SI{10.5}{\micro\kelvin}; also shown are the separate contributions due to states where $n_x,n_y,n_z$ are all even (peaked states) and states where two of $n_x,n_y,n_z$ are odd (nodal states). The inset shows the slow decay of nodal states. (b) 1D slice of the thermal ensemble density along the weak-trapping axis of the tweezer at different times since illumination with photoassociation light started. Simulations used trapping frequencies $(\omega_x,\omega_y,\omega_z) = 2\pi \times (93.0,93.93,20.0)$~kHz, consistent with our experimental setup, and $\Gamma = $~\SI{2.802}{\tera\hertz}. }
	\label{fig:PA_sim_evolution}
\end{figure}

Our simulations qualitatively capture the two-timescale behavior seen in the experiment. As shown in Fig.~\ref{fig:PA_sim_evolution}(a), the fast timescale dynamics are primarily due to the decay of peaked states, whereas the decay of nodal states occurs on a slower timescale (see inset). Figure~\ref{fig:PA_sim_evolution}(b) shows the density of the thermal ensemble after applying different durations of photoassociation. The peak at zero atom-atom separation in the $t=0$ plot is due to the bosonic enhancement of correlations in a thermal cloud. 
We observe a fast depletion of the density around zero atom-atom separation when evolving via Eq.~(\ref{Eq:SE}), with the resultant density dip indicating that there is a low probability of both atoms being found at the same position. 
This rapid transition from a positively pair-correlated ensemble to an anti-correlated ensemble is reflected by the population dynamics [Fig.~\ref{fig:PA_sim_evolution}(a)], confirming that states peaked at $\mathbf{r}=0$ photoassociate fast, while nodal states remain. As the system approaches the single-state limit applicable to groundstate-cooled tweezer experiments~\cite{Kaufman_2012}, the population of nodal states approaches zero and our simulations predict that photoassociation is dominated by the groundstate dynamics and follows a fast decay.

The photoassociation of nodal states is considerably slower in the simulation than in the experiment. A significant contribution to this discrepancy is the lower simulation temperature~\cite{footnote1}. Increasing the temperature rapidly increases the slow-decay rate relative to the fast-decay rate and furthermore increases the fraction of nodal (slow decaying) states~\cite{Suppl}. Increasing the Condon radius also increases the slow rate relative to the fast rate~\cite{Suppl}. The overall agreement between simulated and experimental timescales might also be improved with a different choice of $\Gamma$. Regarding the experiment, we have confirmed that population is not redistributed from nodal to peaked states while the atoms are held in the chopped tweezer~\cite{Suppl}. 
However, any technical imperfection that breaks the reflection symmetry of the tweezer may increase the rate of the slow-decaying states in the experiment.

The experiments also show a higher proportion of fast-decaying states than the simulation. Again, the difference in temperature can contribute to this, since highly-excited nodal states can also decay fast. Additionally, our preparation mechanism might not create a thermal equilibrium distribution. 
Finally, atom-atom interactions change the proportion of peaked to nodal states, although an estimate of the groundstate energy shift indicates that this effect is small.

\emph{Rate coefficients.---} In many-atom ensembles, the single rate coefficient $K_2$ governing the photoassociation dynamics reaches its highest, unitarity-limited value for photoassociation light at the saturation intensity~\cite{Bohn_Julienne_1999, Ospelkaus_2010, Dutta_2014}. For photoassociation of two indistinguishable particles with a maximum photoassociation cross section of $\sigma = \lambda_\text{dB}^2 / (2 \pi)$, the unitarity-limited rate coefficient is~\cite{Kraft_2005}:
\begin{equation}
K_2^{\mathrm{unitarity}} = \sqrt{ 8 \pi \hbar^4 / (\mu^3 k_B T) }.
\label{eq:rate_coefficient_uni}
\end{equation}
This highest achievable rate coefficient forms a fundamental limit that exists for every scattering process, however it has only been investigated in many-atom systems so far.

We determine $K_2$ from the experimentally-observed pair-loss rate $\gamma_2$ of our trapped atom pair~\cite{Suppl, Roberts_2000}:
\begin{equation}
    K_{2} = \frac{\gamma_2}{\int d\mathbf{r}\,[n(\mathbf{r})]^2}, 
\label{eq:rate_coefficient_PA}
\end{equation}
where $n(\textbf{r})$ is the normalized thermal density in the tweezer~\cite{Grimm_2000, Suppl}. Since we model photoassociation in our system with two timescale decay, there are two rate coefficients.
Our experiments are performed close to the saturation intensity in order to compare to the corresponding unitarity-limited rate coefficient [Eq.~(\ref{eq:rate_coefficient_uni})].

\begin{figure}
  \includegraphics[width=\figWidth mm]{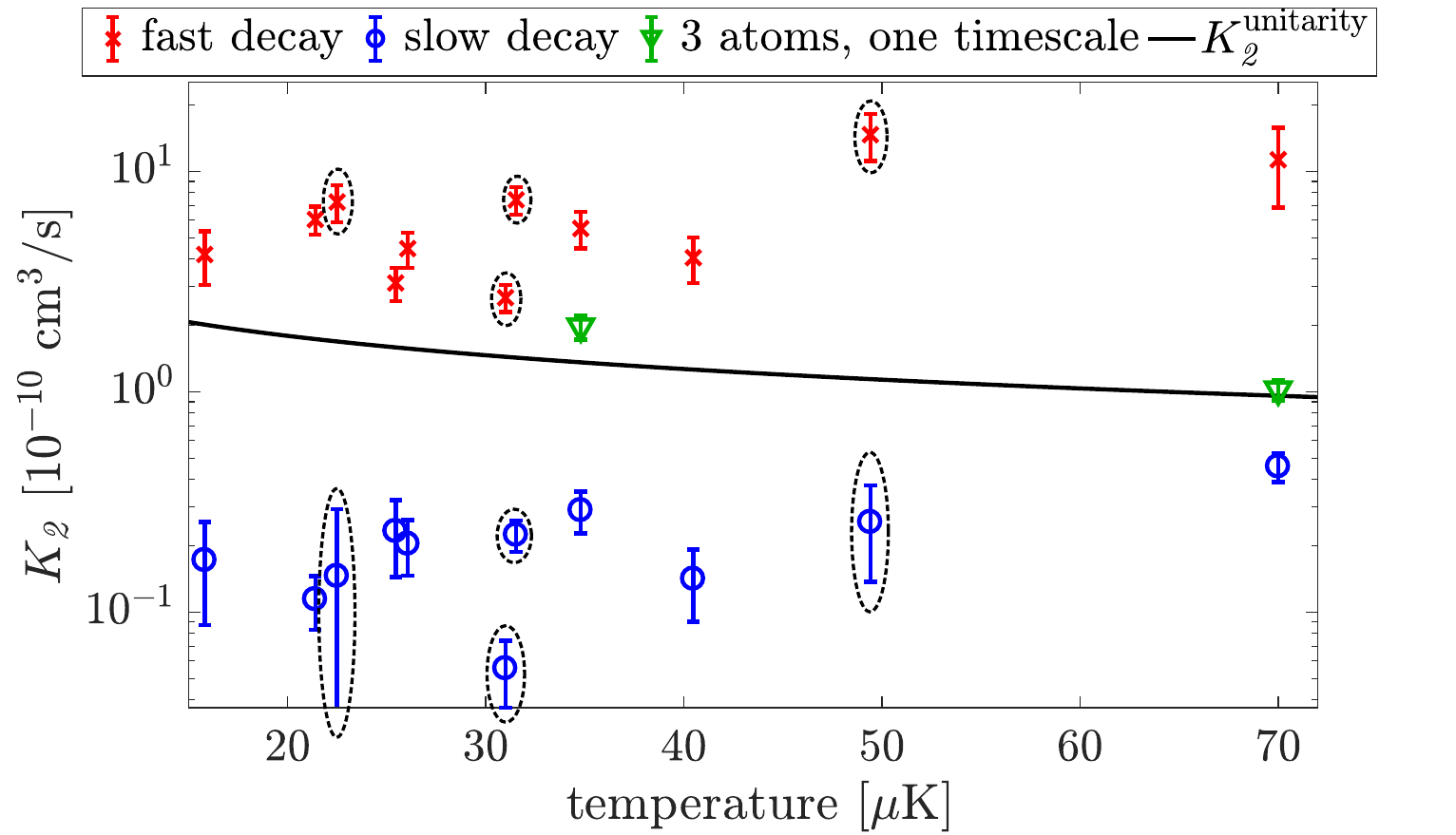}
	\caption{Photoassociation rate coefficients with exactly two atoms in the optical tweezer for fast-decaying populations (ascribed primarily to peaked states) and slow-decaying populations (ascribed to nodal states), compared to the unitarity-limited rate coefficient [Eq.~(\ref{eq:rate_coefficient_uni})]. Points in dashed circles show data at a light intensity of \SI{230}{\watt\per\square\centi\metre} with other points corresponding to \SI{580}{\watt\per\square\centi\metre}. Green triangles show the one-timescale rate coefficient in an optical tweezer with three atoms. The mean mode occupations of these experiments range from $\bar{n}_r \sim 4 - 20$ for the radial trapping directions and $\bar{n}_z \sim 30 - 100$ for the axial trapping direction.}
	\label{fig:K_over_T}
\end{figure}

Figure~\ref{fig:K_over_T} shows the photoassociation rate coefficients of fast and slow decaying populations in our two-atom experiments around the saturation intensity, and compares to the many-atom unitarity limit. The fast decay exceeds $K_2^{\mathrm{unitarity}}$, whereas the rate coefficient for slow decay remains far below it. Recall that the many-atom unitarity-limited rate assumes that collisions happen randomly and depend on the average ensemble density. A plausible explanation for the fast decay exceeding the unitarity limit could be that peaked states have a higher probability of being at zero relative position than expected from the ensemble-averaged density. Note that in Fig.~\ref{fig:K_over_T}, a higher temperature does not imply a higher occupation of the tweezer states; in our experiments, atoms are prepared with similar starting temperature and different temperatures are achieved by adiabatically expanding the tweezer.

\emph{Photoassociation dynamics for three atoms.---} As argued above, the two-timescale decay in the two-atom system is caused by pair correlation of the two atoms. Our setup allows the addition of a third atom to the tweezer, which we used to investigate photoassociation in a three-atom system and determine how the addition of an extra atom influences the dynamics.
Figure~\ref{fig:PA_all_evolution}(b) shows the evolution of the tweezer population starting with three atoms at \SI{35}{\micro\kelvin}. We observe good agreement with a single decay rate above our detection limit of 1.6\% and a reduced $\chi^2$~test shows no improvement from the two-timescale model. The error bar of the point at 3.8\,ms does overlap with the one-timescale fit (this is outside the frame of our logarithmic plot). Additional measurements at photoassociation times greater than \SI{4}{\milli\second} give three-atom survival at our detection sensitivity limit, making them consistent with vanishing survival probability.
The green triangles in Fig.~\ref{fig:K_over_T}, showing the single rate coefficient obtained with three atoms, are in good agreement with the many-atom unitarity-limited rate coefficient. Although we are reluctant to draw firm conclusions, these results suggest that the photoassociation dynamics take a substantial step towards the many-atom behavior when an additional atom is added to the tweezer.

\emph{Conclusions and Outlook.---} Photoassociation of single atoms is a promising path for creating precisely tailored single molecules not accessible through conventional chemistry. Understanding the photoassociation dynamics of single molecule formation is vital to the future controlled synthesis of more complex molecules. We showed the first measurement of the quantum dynamics of exactly two atoms undergoing photoassociation in an optical tweezer. We observed two rate coefficients which are caused by atom-pair correlations, as confirmed by numerical simulation of the two-atom system. In contrast, the photoassociation dynamics of three trapped atoms seem closer to that of many-atom ensembles. An interesting future work would be a more systematic investigation into how the many-atom dynamics emerge. 
Our results show that this state-dependent photoassociation could be used as a new tool for the production or detection of atom-pair correlations in future experiments.

\begin{acknowledgments}
\emph{Acknowledgements.---} We acknowledge useful discussions with and comments from Joachim Brand and Eite Tiesinga. This work was supported by the Marsden Fund Council from Government funding, administered by the Royal Society of New Zealand (Contract No. UOO1835). SSS was supported by an Australian Research Council Discovery Early Career Researcher Award (DECRA), Project No. DE200100495. This research was undertaken with the assistance of resources and services from the National Computational Infrastructure (NCI), which is supported by the Australian Government.
\end{acknowledgments}

\nocite{Chernick_Bootstrap}

\providecommand{\noopsort}[1]{}\providecommand{\singleletter}[1]{#1}%
%


\renewcommand{\thefigure}{S\arabic{figure}}
\setcounter{figure}{0}
\renewcommand{\theequation}{S\arabic{equation}}
\setcounter{equation}{0}

\newcommand\mytablewidth{0.062}
\newcommand\mytablewidthl{0.065}

\newpage
\onecolumngrid
\begin{centering}
\large{
\textbf{Supplemental Material for: Pair Correlations and Photoassociation Dynamics of Two Atoms in an Optical Tweezer}
}\\
\vspace{1em}\normalsize{
M. Weyland,$^\textrm{1, 2}$ S. S. Szigeti,$^\textrm{3}$ R. A. B. Hobbs,$^\textrm{1, 2}$ P. Ruksasakchai,$^\textrm{1, 2}$ L. Sanchez,$^\textrm{1, 2}$ and M. F. Andersen$^\textrm{1, 2}$\\
\vspace{1ex}
$^\textrm{1}$\textit{The Dodd-Walls Centre for Photonic and Quantum Technologies, University of Otago, Dunedin, New Zealand}\\
$^\textrm{2}$\textit{Department of Physics, University of Otago, Dunedin, New Zealand}\\
$^\textrm{3}$\textit{Department of Quantum Science, Research School of Physics,\\
The Australian National University, Canberra 2601, Australia}
}\\
\vspace{2em}
\end{centering}
\twocolumngrid

\section{Determination of optical tweezer's populations}
\begin{figure}[t!]
  \includegraphics[width=\figWidth mm]{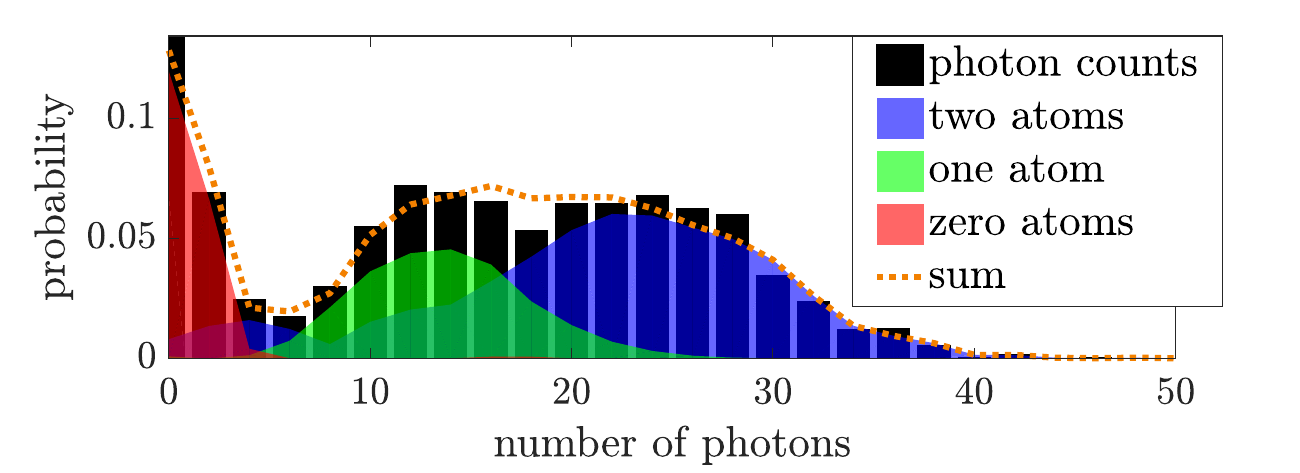} 
	\caption{Example of a photon distribution obtained from 3600 iterations of an experiment which results in a mix of zero-, one- and two-atom survivals (black bars). The fitted contributions to this photon distribution from zero-, one- and two-atom survival are shown as the shaded areas. The photon count distributions for the pure zero-, one- and two-atom population were obtained from calibration measurements. \label{fig:SPCM}}
\end{figure}

When reconstructing the probability of finding a certain number of atoms in the tweezer after photoassociation, as shown in Fig.~2 of the main text, we first record calibration photon count distributions by capturing atoms without exposure to photoassociation light. We obtain the initial atom number from an EMCCD image before merging the traps and sort cases with zero-, one-, and two-atom initial loading. We perform these calibration measurements several thousand times to obtain photon count distributions for each possible number of atoms in the tweezer. These measurements contain 3.5\% single-atom loss from the merging process, which can be identified in the photon count histogram of one-atom loading cases as realizations with low photon counts (see ``zero atoms" in Fig.~\ref{fig:SPCM}). We correct the calibration photon count distributions of one-atom and two-atom loading cases for this loss and fit a combination of the resulting calibration distributions to the experimental photon count distributions, which we obtain from photoassociation measurements, as shown in Fig.~\ref{fig:SPCM}.

Since the parameter space of the fit is restricted to two free parameters between 0\% and 100\% that correspond to the probability of finding zero or one atom, we can ensure that our fitting algorithm finds the global minimum of the squared residuals by probing the whole parameter space with a 5\%-step size, followed by probing the area around the global minimum with a 0.25\%-step size. 
We create 1000 bootstrap resampling data sets from this initial fit to obtain the final parameters and their standard deviations~[36]. 
To analyze experiments with three atoms, we extend the described process to include a three-atom photon count distribution.

\section{Fit of photoassociation dynamics}
The dynamics of the photoassociation process can be described by a set of coupled rate equations. In the specific case of two atoms in an optical tweezer, we describe the evolution of the populations by
\begin{equation} 
\label{eq:pop_evolution_2A_2t}
\left( \begin{array}{c} \dot{p_0} \\ \dot{p_1} \\ \dot{p^n_2} \\ \dot{p^p_2} \end{array} \right) = 
 \begin{pmatrix} 0 & \gamma_1 & \gamma^n_2 & \gamma^p_2 \\ 0 & -\gamma_1 & 2\gamma_1 & 2\gamma_1 \\ 0 & 0 & -\gamma^n_2-2\gamma_1 & 0 \\ 0 & 0 & 0 & -\gamma^p_2-2\gamma_1 \end{pmatrix} 
 \left( \begin{array}{c} p_0 \\ p_1 \\ p^n_2 \\ p^p_2 \end{array} \right),
\end{equation}
where $p_i(t)$ is the probability of finding $i$ atoms in the tweezer after we expose it to photoassociation light for a specific time $t$. Single-atom loss occurs with a rate $\gamma_1$ due to collisions of the atoms with background gas in the vacuum chamber. 
This single atom loss is around one order of magnitude slower than the slow two-atom loss at rate $\gamma_2^n$. The two different probabilities for the two-atom population describe the fast decaying tweezer states which are primarily ascribed to relative-motional wavefunctions with a peak at $r=0$ ($p^p_2$) and slow decaying wavefunctions which are primarily ascribed to states with a node at zero separation ($p^n_2$). Our measured probability for the total two-atom population is thus $p_2 = p^n_2 + p^p_2$. In our analysis we compare this two-timescale model with the one-timescale model for which the second two-atom population is removed. 
When analyzing the evolution of a three-atom system, the equation is extended to include a probability for three-atom population but no three-atom loss. In this extension of the equation, single-atom loss starting from a three-atom population is assumed to populate both two-atom populations ($p^n_2$ and $p^p_2$) equally.
The initial populations of fast and slow decaying states depend on temperature and trapping frequencies. They are left as a fitting parameter.
All measured rates of photoassociation and single-atom loss can be found in Table~\ref{tab:gamma_data}. For comparison, the longest photoassociation time used in the experiments is 120\,ms. Single-atom loss $\gamma_1$ is determined from the photoassociation time, not the total time of holding the atoms in the chopped tweezer. Since photoassociation light is only running for a fraction of the time while the tweezer is turned off, this corresponds to 8.4\% of the actual hold time in the chopped tweezers.

\begin{table}
\caption{Rates for single-atom loss ($\gamma_1$) and photoassociation of slow-decaying states ($\gamma_2^n$; ascribed to nodal states) as well as fast-decaying states ($\gamma_2^p$; primarily ascribed to peaked states). Measurements from photoassociation with three atoms show results from one-timescale fits. Errors shown are $1\sigma$ confidence intervals of the fits.}
\begin{tabular}{|l|p{\mytablewidthl\textwidth}p{\mytablewidthl\textwidth}|p{\mytablewidth\textwidth}p{\mytablewidth\textwidth}|p{\mytablewidth\textwidth}p{\mytablewidth\textwidth}| }
\hline
$T$        & $\gamma_1$ & $\Delta\gamma_1$ & $\gamma_2^n$ & $\Delta\gamma_2^n$ & $\gamma_2^p$ & $\Delta\gamma_2^p$ \\
{[$\mu$K]} & [$s^{-1}$] & [$s^{-1}$]       & [$s^{-1}$]   & [$s^{-1}$]         & [$s^{-1}$]   & [$s^{-1}$]       \\
\hline
15.8     & 1$\cdot 10^1$    & 1$\cdot 10^1$   & 4$\cdot 10^1$    & 2$\cdot 10^1$  & 1.1$\cdot 10^3$ & 3$\cdot 10^2$              \\
19.2     & 0                & 1$\cdot 10^1$   & 1.3$\cdot 10^2$  & 3$\cdot 10^1$  & 1.1$\cdot 10^3$ & 2$\cdot 10^2$                \\
21.4     & 3                & 4               & 1.9$\cdot 10^1$  & 5              & 1.0$\cdot 10^3$ & 1$\cdot 10^2$                \\
22.5     & 6                & 6               & 1$\cdot 10^1$    & 1$\cdot 10^1$  & 5$\cdot 10^2$   & 1$\cdot 10^2$                \\
25.5     & 1$\cdot 10^1$    & 3$\cdot 10^1$   & 1.6$\cdot 10^2$  & 6$\cdot 10^1$  & 2.2$\cdot 10^3$ & 3$\cdot 10^2$                \\
26.1     & 1$\cdot 10^1$    & 1$\cdot 10^1$   & 6$\cdot 10^1$    & 2$\cdot 10^1$  & 1.4$\cdot 10^3$ & 2$\cdot 10^2$                \\
31.0     & 5                & 7               & 2.1$\cdot 10^1$  & 7              & 1.0$\cdot 10^3$ & 1$\cdot 10^2$                \\
31.5     & 1                & 3               & 2.9$\cdot 10^1$  & 4              & 9$\cdot 10^2$   & 1$\cdot 10^2$                \\
31.8     & 0                & 3               & 1.1$\cdot 10^1$  & 4              & 3.4$\cdot 10^2$ & 4$\cdot 10^1$               \\
34.8     & 7                & 6               & 9$\cdot 10^1$    & 2$\cdot 10^1$  & 1.7$\cdot 10^3$ & 3$\cdot 10^2$                \\
40.5     & 3  $\cdot 10^1$  & 2$\cdot 10^1$   & 7$\cdot 10^1$    & 2$\cdot 10^1$  & 1.9$\cdot 10^3$ & 4$\cdot 10^2$                \\
49.4     & 0                & 3               & 6                & 3              & 3.4$\cdot 10^2$ & 8$\cdot 10^1$                 \\
70.0     & 1.6$\cdot 10^1$  & 5               & 5.0$\cdot 10^1$  & 7              & 1.2$\cdot 10^3$ & 5$\cdot 10^2$                \\
\hline
\hline
$T$       & $\gamma_1$ & $\Delta\gamma_1$ &\multicolumn{2}{c|}{\multirow{2}{*}{3 atoms}}& $\gamma_2$ & $\Delta\gamma_2$ \\
{[$\mu$K]} & [$s^{-1}$] & [$s^{-1}$]     &\multicolumn{2}{c|}{}              & [$s^{-1}$]   & [$s^{-1}$]       \\
\hline
34.8     & 4.7              & 9$\cdot 10^{-1}$    & ---          & ---                & 6.2$\cdot 10^2$ & 7$\cdot 10^1$                  \\
70.0     & 1.6$\cdot 10^1$  & 2                   & ---          & ---                & 1.1$\cdot 10^2$ & 1$\cdot 10^1$                  \\
\hline
\end{tabular}
\label{tab:gamma_data}
\end{table}

\section{Photoassociation parameters}
\begin{figure}
  \includegraphics[width=\figWidth mm]{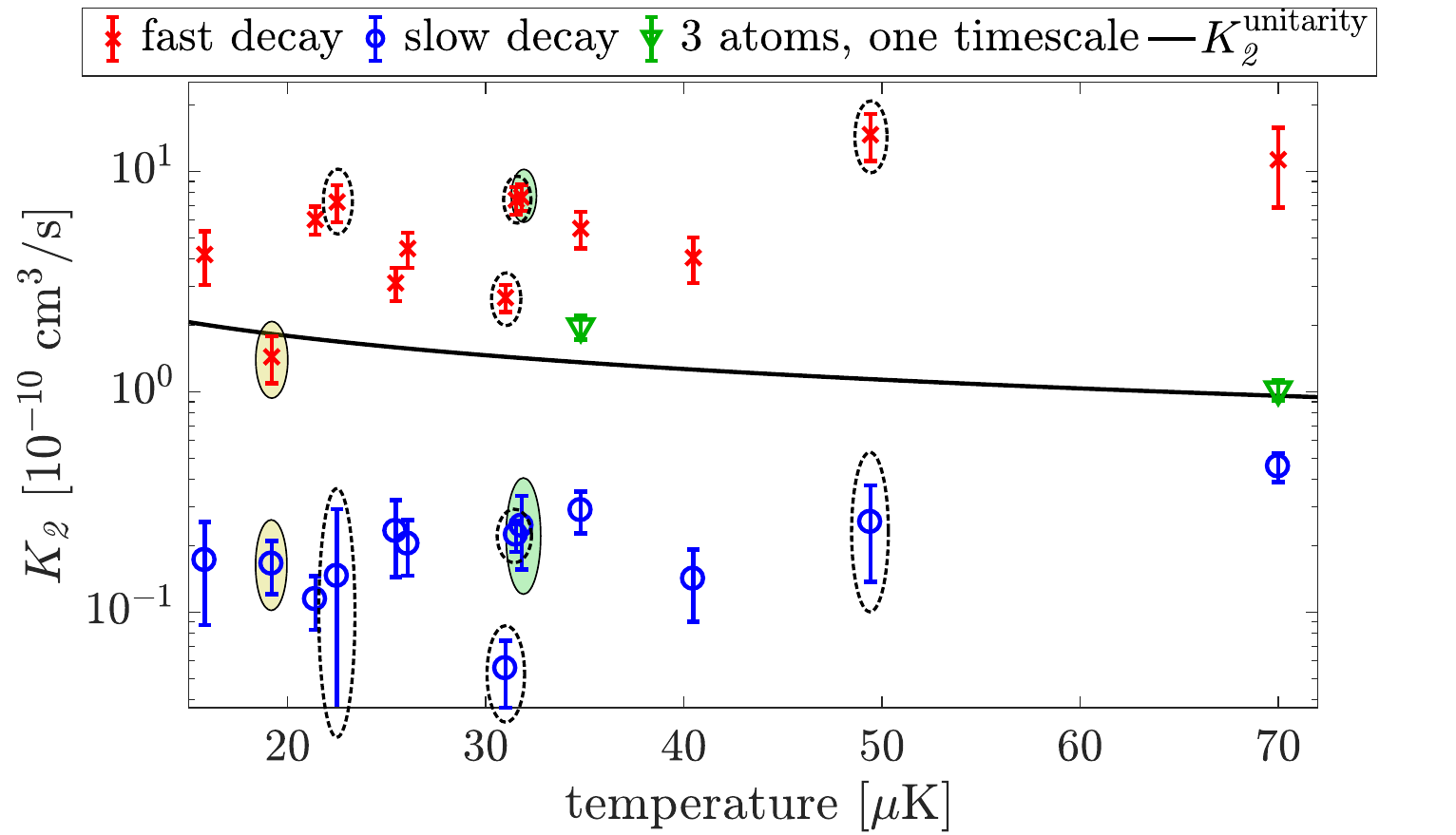}
	\caption{Photoassociation rate coefficients for fast-decaying populations (red crosses; ascribed primarily to peaked states) and slow-decaying populations (blue circles; ascribed to nodal states), compared to the many-atom unitarity-limited rate coefficient. Specially marked measurements are described in the Supplemental text. \label{fig:K_over_T_all}}
\end{figure}

Most measurements were performed at a peak laser intensity of \SI{580}{\watt\per\square\centi\metre}. Some additional measurements were performed with different parameters and are shown in Fig.~\ref{fig:K_over_T_all}. The points in dashed circles were taken at a peak intensity of \SI{230}{\watt\per\square\centi\metre}, closer to the intensity which showed the highest rate coefficient. The solid-encircled points were taken without chopping the tweezer and with a photoassociation light intensity of \SI{580}{\watt\per\square\centi\metre} (yellow filled) and \SI{21}{\watt\per\square\centi\metre} (green filled). These results rule out our experimental procedure of chopping tweezer and photoassociation light as a possible source of the observed two-timescale dynamics. 

\begin{figure}
  \includegraphics[width=\figWidth mm]{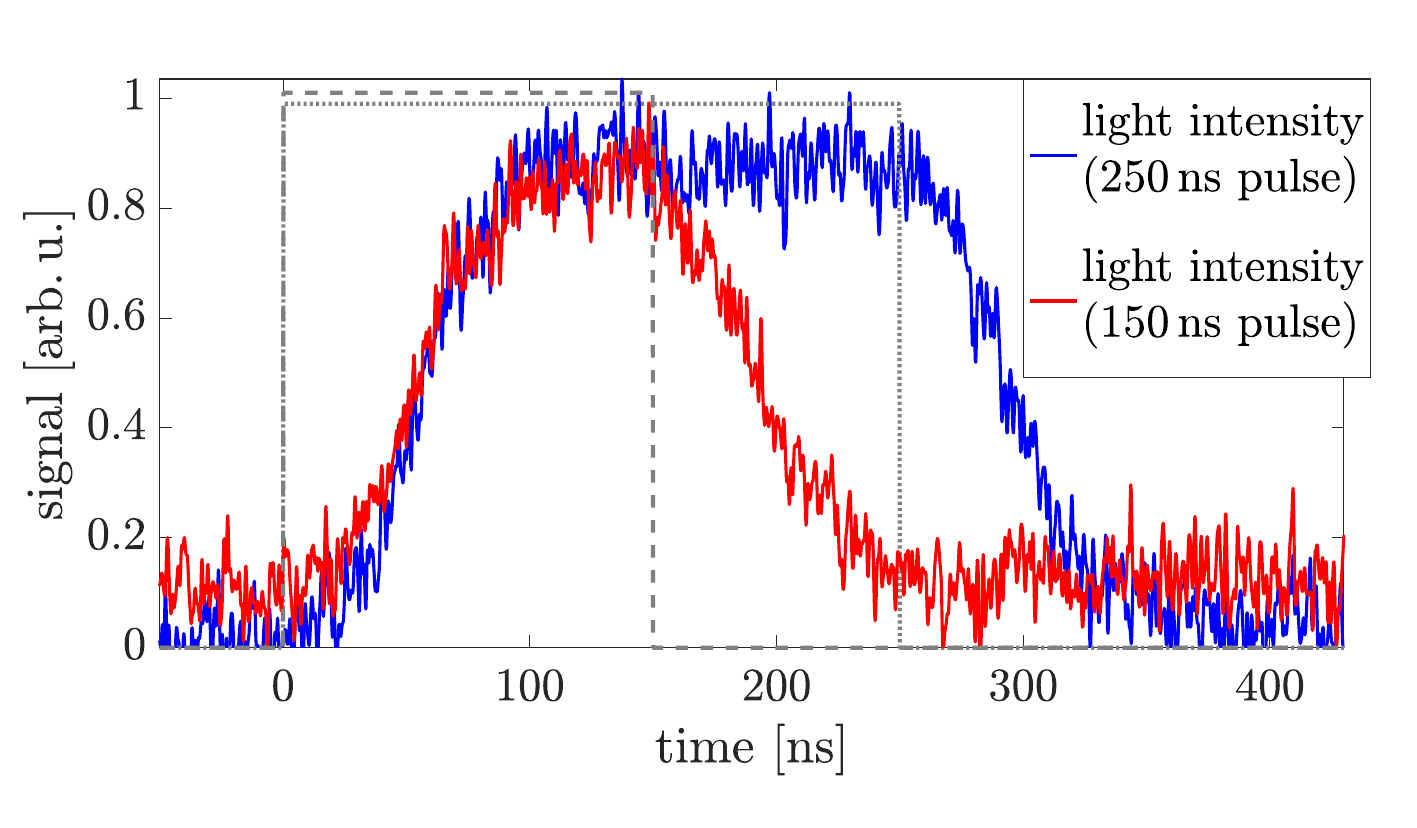}
	\caption{Shape of the actual photoassociation pulses (\SI{150}{\nano\second}) and of a longer photoassociation pulse (\SI{250}{\nano\second}) for comparison. The dashed and dotted lines show the pulse shapes for triggering the \SI{150}{\nano\second} and \SI{250}{\nano\second} photoassociation pulses. They have been shifted in time by a common value to have their falling edge coincide with the beginning of the fall time of the photoassociation pulses.\label{fig:PA_pulse_shape}}
\end{figure}

We measured the pulse shape using a fast photo-diode (HFD3180-203, \SI{80}{\pico\second} rise/fall time) and a \SI{500}{\mega\hertz} oscilloscope. The observed pulse shape is shown in Fig.~\ref{fig:PA_pulse_shape}. In addition to the \SI{150}{\nano\second} pulse used in our measurements we also plot a \SI{250}{\nano\second} pulse which shows that the maximum intensity is reached in the experiment. However, we observe significant rise and fall times. The maximum intensity is maintained for only \SI{50}{\nano\second} and above 50\% intensity for \SI{140}{\nano\second}. In our analysis we use this 50\%-threshold to define our pulse length and the resulting photoassociation time.

We prepared atoms in the \mbox{$|$F\,=\,2,\,m\,=\,-2$\rangle$} state in two tweezers with an efficiency of 98\%. The m-state gets conserved during the merging process with an efficiency of 80\%. We performed one measurement with an improved merging process (data point at \SI{49.4}{\micro\kelvin} in Fig.~4 and Fig.~\ref{fig:K_over_T_all}), which retained the prepared m-state with an efficiency of 98\%. Using this improved merging process did not alter the observed photoassociation dynamics.

\section{Control measurements}
We performed control measurements for each of our photoassociation measurements in which we chopped the tweezers for the same duration but blocked the photoassociation light. We fitted the population as a function of chopping time and found no two-body loss on the timescale of our experiments (\SI{1.5}{\second} chopping time, corresponding to \SI{120}{\milli\second} of photoassociation). 

Additionally, we checked for mixing of peaked and nodal states in our tweezer during the experiment using the following sequence:
\begin{enumerate}[topsep=1ex,itemsep=-1ex,partopsep=1ex,parsep=1ex]
\item first instance of photoassociation light for 28\,ms,
\item variable wait time while chopping the tweezer,
\item second instance of photoassociation light for 28\,ms.
\end{enumerate}
Each instance of photoassociation is enough to deplete the fast-decaying states. If states are re-thermalized on the timescale of the experiment, then the probability for photoassociation should increase with the wait time as fast-decaying states get repopulated. However, even for wait times as long as the longest experimental duration we observe no change in photoassociation probability. Re-thermalization is therefore an unlikely candidate to explain the quantitative discrepancy between the slow decay rate in experiment and simulation.

\section{Statistical analysis of fits}
We assess the validity of our two-timescale model when fitting it to 19 different sets of photoassociation measurements.
These experiments were performed using different atom temperatures and photoassociation light powers as well as changing the preparation of the atoms from the \mbox{$|$F\,=\,2,\,m\,=\,-2$\rangle$} state to a mix of all five m-states. A $\chi^2$-test~[29] rejects the one-timescale model for 17 of those measurements at the $5\%$ level of significance, whereas the two-timescale model is rejected for only 2 measurements. 
 Using the two-timescale model improves the reduced $\chi^2$-values by a factor of 7 compared to the one-timescale model.

\section{Calculation of rate coefficient}
Equation~(6) of the main text shows the relation between the pair-loss rate $\gamma_2$ in the experiment and the underlying rate coefficient $K_{2}$. It requires knowledge of the thermal density of an atom in the optical tweezer, which is given by~[35] 
\begin{equation}
    n(\mathbf{r}) = n_0\,\mathrm{exp}\left(\frac{-(U(\mathbf{r}) - U_0)}{k_B T}\right),
    \label{eq:density_generic}
\end{equation}
where $n_0$ is the peak density and $U(\mathbf{r})$ is the trapping potential with depth $U_0$.
At the temperatures used in our experiments, the atoms are in low trap states. Therefore, the Gaussian trap can be approximated as a harmonic potential, giving
\begin{equation}
     n(\mathbf{r}) = n_0\,\mathrm{exp}\left(\frac{-m \omega_r^2 (x^2 +y^2)}{2 k_B T}\right)\,\mathrm{exp}\left(\frac{-m \omega_z^2 z^2}{2 k_B T}\right).
    \label{eq:density_our_tweezers}
\end{equation}
We measured the radial frequency $\omega_r$ using Raman side-band spectroscopy and obtained the axial frequency $\omega_z$ from the previously measured ratio between radial and axial frequencies in our tweezer~[26]. 

\section{Determination of saturation intensity}
The expected saturation intensity can be found from the resonance width as a function of photoassociation intensity~[32,~33]. 
We measured the width of the photoassociation resonance in an intensity range from \SI{12}{\watt\per\square\centi\metre} (\SI{0.1}{\milli\watt} beam power) to \SI{2300}{\watt\per\square\centi\metre} (\SI{200}{\milli\watt} beam power) and used a weighted linear fit to obtain a linewidth of \SI[separate-uncertainty = true]{12(6)}{\mega\hertz} at zero intensity, which serves as an upper bound for the natural linewidth. We find the saturation intensity - where the observed linewidth is twice the natural linewidth -  to be \SI[separate-uncertainty = true]{200(100)}{\watt\per\square\centi\metre}. 
Although this method gives large errors, it does not rely on any rate measurement and the expected saturation intensity is in the range where we performed our experiments.

\section{Details of numerical simulation}
We numerically solve Eq.~(3) of the main text by expanding $\psi(\textbf{r})$ on a finite basis of even-parity eigenstates of $H_\text{rel}(\textbf{r})$: $\psi(\textbf{r},t) = \sum_{\varepsilon_\textbf{n} \leq E_\text{cut}} c_n(t) \phi_\textbf{n}(\textbf{r})$, where $\textbf{n} = (n_x,n_y,n_z)$ and the sum is over all even-parity eigenstates with energy $\varepsilon_\textbf{n} = \sum_{i=x,y,z} \hbar \omega_i (n_i+\tfrac{1}{2})$ less than some energy cutoff $E_\text{cut}$. 
We assume our initial condition is a thermal distribution of even eigenstates of $H_\text{rel}(\textbf{r})$. That is, in any given experiment $\psi(\textbf{r},0) = \phi_{\textbf{n}_0}(\textbf{r})$ with Boltzmann probability $\mathcal{P}_{\textbf{n}_0} = \exp(-\beta \varepsilon_{\textbf{n}_0}) / \mathcal{Z}$,
where $\beta = 1 / k_\textrm{B} T$ and the partition function $\mathcal{Z}$ has an analytic expression (see for example Supplementary Note 5 of [21]): 
\begin{align}
	\mathcal{Z}     &= \mathcal{Z}_x^\text{even} \mathcal{Z}_y^\text{even} \mathcal{Z}_z^\text{even} + \mathcal{Z}_x^\text{odd} \mathcal{Z}_y^\text{odd} \mathcal{Z}_z^\text{even} \notag \\
	                &+ \mathcal{Z}_x^\text{odd} \mathcal{Z}_y^\text{even} \mathcal{Z}_z^\text{odd} + \mathcal{Z}_x^\text{even} \mathcal{Z}_y^\text{odd} \mathcal{Z}_z^\text{odd},
\end{align}
where
\begin{align}
	\mathcal{Z}_i^\text{even} 	&= \sum_{m_i=0}^\infty e^{- \beta \hbar \omega_i (2 m_i + \frac{1}{2})} = \frac{e^{-\beta \hbar \omega_i/2}}{1 - e^{-2\beta \hbar \omega_i}}, \\
	\mathcal{Z}_i^\text{odd} 	&= \sum_{m_i=0}^\infty e^{- \beta \hbar \omega_i [ (2m_i+1) + \frac{1}{2}]} = \frac{e^{-3\beta \hbar \omega_i/2}}{1 - e^{-2\beta \hbar \omega_i}}.
\end{align}
We choose $E_\text{cut}$ such that both $\sum_{\varepsilon_\textbf{n} \leq E_\text{cut}}\mathcal{P}_\textbf{n} \gtrsim 0.996$ and coupling to the highest-energy, sparsely-occupied modes is negligible. For the computational resources at our disposal, these conditions limit our calculations to temperatures no greater than $\SI{10.5}{\micro \kelvin}$ -- roughly half the lowest temperature investigated experimentally.

In this basis, evolution under Eq.~(3) corresponds to
\begin{align}
    i \hbar \dot{c}_\textbf{n} = \varepsilon_\textbf{n} c_\textbf{n} -i \sum_{\textbf{m} \leq E_\text{cut}} V_{\textbf{n}\textbf{m}} c_\textbf{m}, \label{HG_evo}
\end{align}
where
\begin{equation}
    V_{\textbf{n}\textbf{m}} = \int d\textbf{r} \, \phi_\textbf{m}^*(\textbf{r}) V_\text{PA}(\textbf{r}) \phi_\textbf{n}(\textbf{r}).
\end{equation}
In order to efficiently compute the matrix elements $V_{\textbf{n}\textbf{m}}$, we first write the hard-shell potential $V_\text{PA}(\textbf{r})$ (Eq.~(4) of the main text) as the difference of two hard-sphere potentials: $V_\text{PA}(\textbf{r}) = V_\text{sphere}^{r_c}(\textbf{r}) - V_\text{sphere}^{r_c-w}(\textbf{r})$, where
\begin{equation} \label{Eq:hard_sphere_suppl}
    V_\text{sphere}^R(\textbf{r}) = \begin{cases} 
                                \hbar \Gamma, & |\textbf{r}| \leq R, \\
                                0,   & \textrm{otherwise}.
   \end{cases}
\end{equation}
Then $V_{\textbf{n}\textbf{m}} = V_{\textbf{n}\textbf{m}}^{r_c} - V_{\textbf{n}\textbf{m}}^{r_c-w}$, where
\begin{widetext}
\begin{align}
    V_{\textbf{n}\textbf{m}}^R   &= \int d\textbf{r} \, \phi_\textbf{m}^*(\textbf{r})V_\text{sphere}^R(\textbf{r}) \phi_\textbf{n}(\textbf{r}) \notag \\
                    &= \hbar \Gamma \int_{-R}^{R} dx \, \varphi_{m_x}^*(x)\varphi_{n_x}(x) \int_{-\sqrt{R^2 - x^2}}^{\sqrt{R^2 - x^2}} dy \, \varphi_{m_y}^*(y)\varphi_{n_y}(y)\int_{-\sqrt{R^2 - x^2 - y^2}}^{\sqrt{R^2 - x^2-y^2}} dz \, \varphi_{m_z}^*(z)\varphi_{n_z}(z). \label{V_HG}
\end{align}
\end{widetext}
These matrix elements were numerically computed for $R = r_c$ and $R = r_c - w$ using adaptive Hermite-Legendre quadrature methods in Mathematica.

For a given initial condition $\psi(\textbf{r},0) = \phi_{\textbf{n}_0}(\textbf{r})$, solving Eq.~(\ref{HG_evo}) allows us to compute the probability $N_{\textbf{n}_0}(t) = \sum_{\varepsilon_\textbf{n} \leq E_\text{cut}} |c_\textbf{n}(t)|^2$. 
The total population assuming a thermal initial state is given by an incoherent sum over $N_{\textbf{n}_0}(t)$ weighted by the Boltzmann probability $\mathcal{P}_{\textbf{n}_0}$:
\begin{equation}
	P(t) = \sum_{\varepsilon_{\textbf{n}_0} \leq E_\text{cut}} \mathcal{P}_{\textbf{n}_0} N_{\textbf{n}_0}(t).
\end{equation}
This procedure was used to generate the simulation data plotted in Fig.~3 of the main text.

We obtained a significant computational efficiency by exploiting the structure of the hard-sphere / hard shell potential matrix elements. Explicitly, the 1D integrals in Eq.~(\ref{V_HG}) are only non-zero if $n_i$ and $m_i$ are both even or both odd for $i = x,y,z$. Consequently, Eq.~(\ref{HG_evo}) can be reduced to evolution within four decoupled subspaces, corresponding to basis states spanned by even-parity eigenstates $\phi_{\textbf{n}}(\textbf{r})$ where (eee) $n_x, n_y, n_z$ are all even, (eoo) $n_x$ is even and $n_y, n_z$ are odd, (oeo) $n_x, n_z$ odd and $n_y$ even, and (ooe) $n_x, n_y$ odd and $n_z$ even. Physically, only states in the `eee' subspace have a non-zero density at relative co-ordinate $\textbf{r} = 0$. It is therefore states in this subspace that decay the fastest under our photoassociation model and which are predominantly responsible for the fast timescale decay shown in Fig.~2 of the main text.

\section{Dependence of the simulation on temperature and absorbing potential range}
\begin{figure}[t]
\includegraphics[width=\figWidth mm]{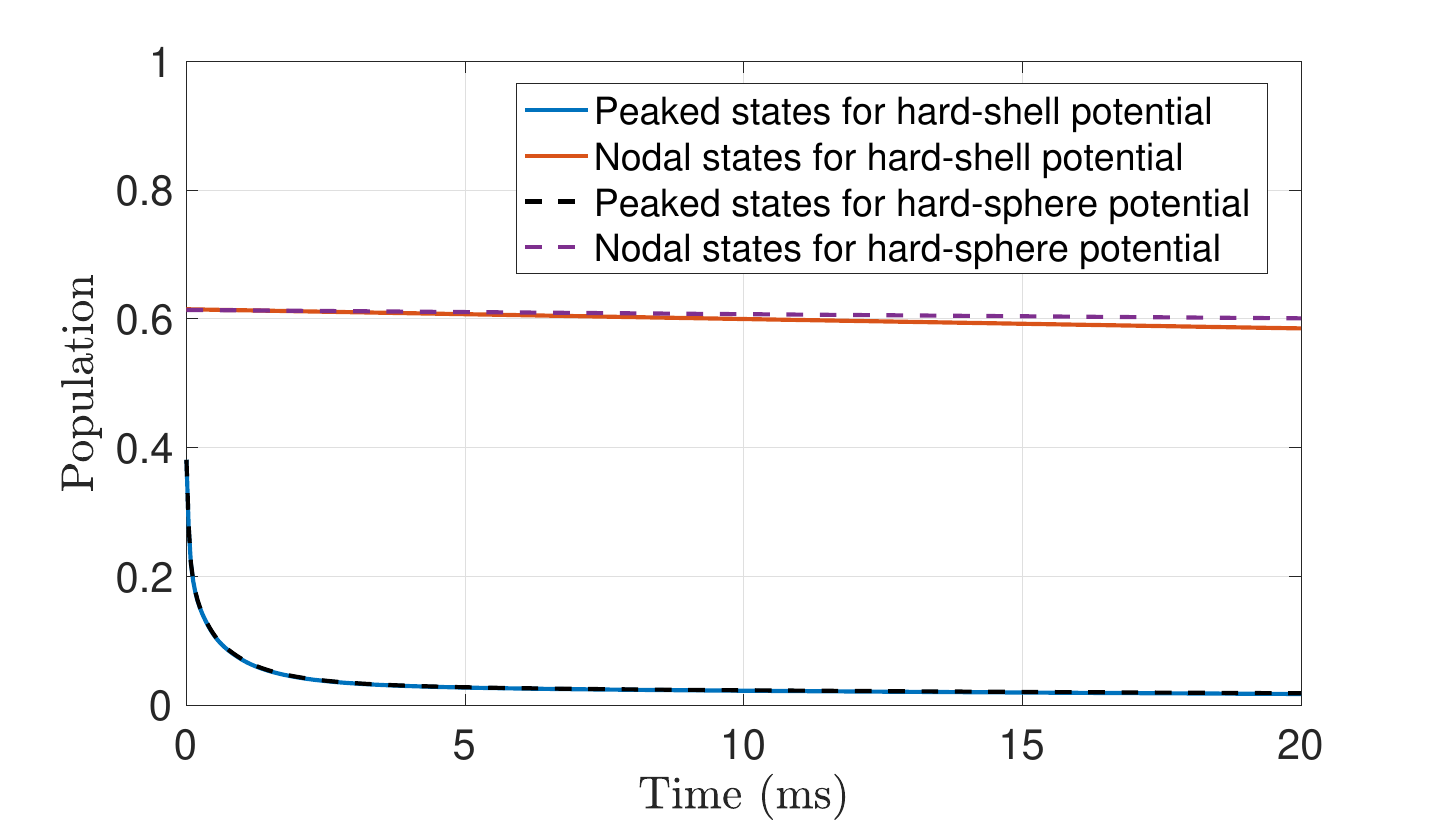}
	\caption{Comparison of simulated photoassociation at 10.5$\mu$K using an absorbing hard-shell potential (solid lines) and an absorbing hard-sphere potential (dashed lines). For a fair comparison, we chose $\Gamma$ such that the integral over $V_\text{PA}$ (i.e. the loss rate coefficient) remained constant. Specifically, the strength of the hard-shell potential was $\Gamma = K / \mathcal{V}_\text{shell}$, whereas it was $\Gamma = K / \mathcal{V}_\text{sphere}$ for the hard-sphere potential, where $K = 2.474 \times 10^{-16}$~m$^3$/s and $\mathcal{V}_\text{sphere} = \frac{4}{3}\pi r_c^3$ and $\mathcal{V}_\text{shell} = \frac{4}{3}\pi [r_c^3 - (r_c-w)^3]$ are the non-zero absorbing volumes of the sphere and shell potentials, respectively.}
	\label{sphere_vs_shell_comparison}
\end{figure}

Here we systematically show how the photoassociation dynamics predicted by our simulation changes as we (1) change the ensemble temperature and (2) increase the non-zero region of the absorbing potential to interatomic distances greater than the Condon radius $r_c = 4.3$nm. For the simulations used in this analysis, we treat loss due to photoassociation as a hard-sphere potential [Eq.~(\ref{Eq:hard_sphere_suppl})]. As shown in Fig.~\ref{sphere_vs_shell_comparison}, this gives very similar results to the absorbing hard-shell potential; specifically, identical dynamics are observed for the fast-decaying peaked states, while nodal states decay around twice as fast in the hard-shell case.

\begin{figure}[t]
	\begin{center}
	\vspace{1em}
  \includegraphics[width=\figWidth mm]{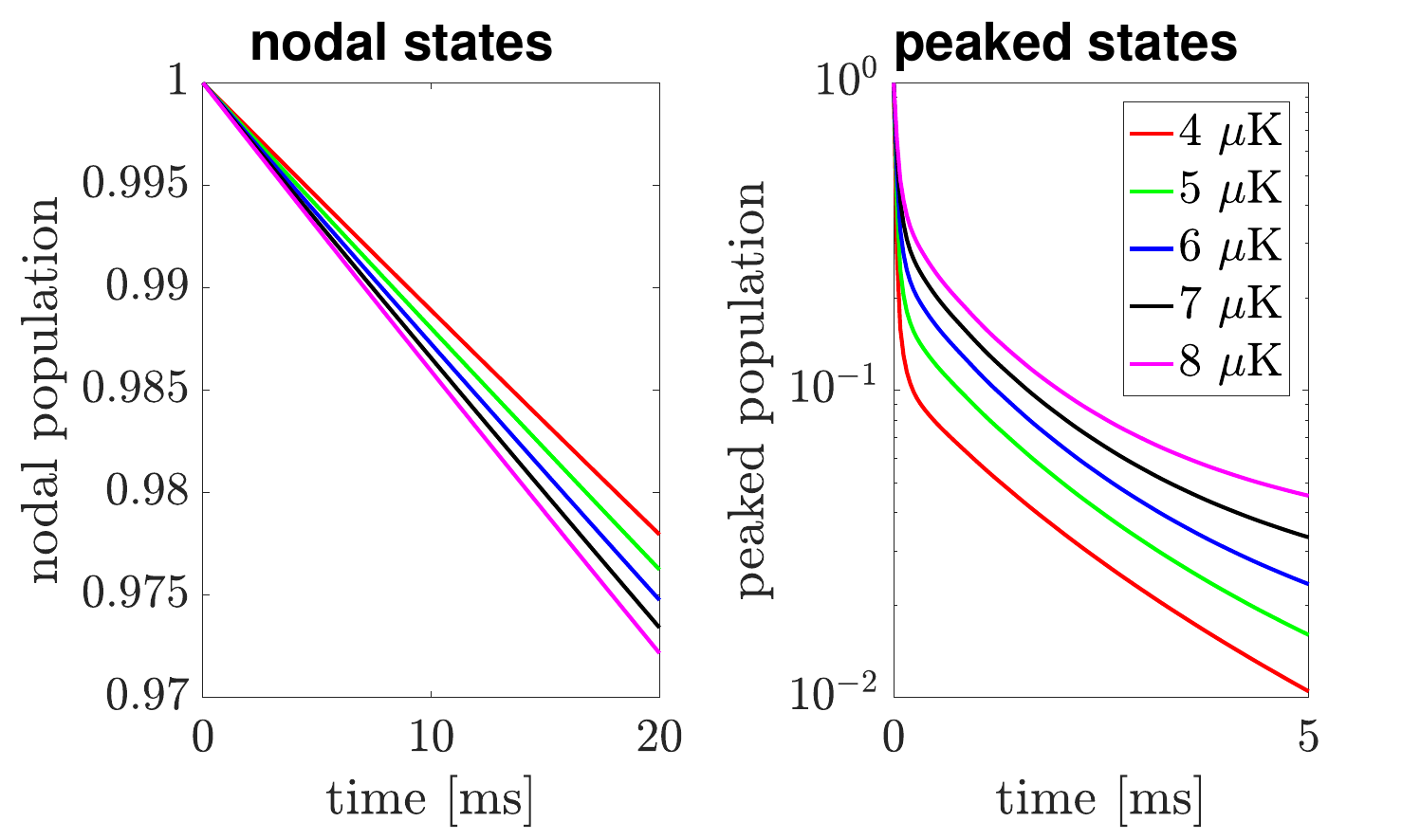}
  \end{center}
	\caption{Change of the photoassociation dynamics at different temperatures, separated by state symmetry. All simulations use a hard sphere potential and the same photoassociation strength of $\Gamma = 1.076$~GHz.}\label{fig:sim_n_p_T}
\end{figure}

Figure~\ref{fig:sim_n_p_T} shows how changes to the temperature affect the decay of the states peaked at $\textbf{r} = 0$ and the nodal states. For all temperatures, note the very different decay timescales of the peaked and nodal states. As temperature increases, the simulated photoassociation dynamics show an increase in the decay rate of the nodal states and a decrease in the decay of the peaked states. The ratio of slow decay (nodal state) timescale to fast decay (peaked state) timescale is highly sensitive to temperature: it drops from 22000 at \SI{4}{\micro\kelvin} to 3600 at \SI{8}{\micro\kelvin}. For comparison, the experimentally-observed ratio between timescales is around 30 at \SI{35}{\micro\kelvin}. This suggests that the lower temperature used in the simulations is a major contribution to the quantitative disagreement between the simulated and experimentally-observed slow decay timescales reported in the main text.

\begin{figure}[t!]
	\begin{center}
     \includegraphics[width=\figWidth mm]{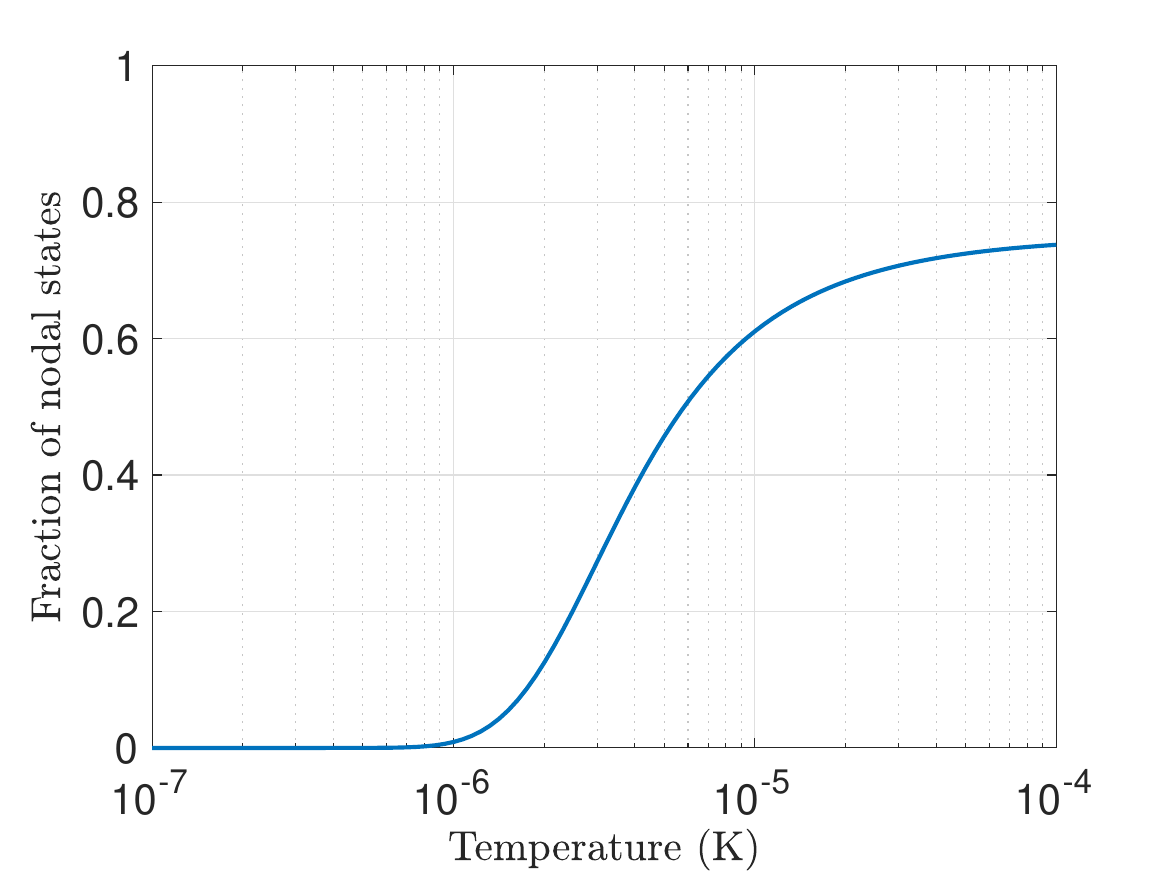}
	\end{center}
   	\caption{Fraction of nodal (even-odd-odd, odd-even-odd, odd-odd-even states) atom-pair states in a non-interacting thermal distribution, assuming a harmonic trapping potential with frequencies $\omega_x = 2\pi \times 93$~kHz, $\omega_y = 1.01 \omega_x$, and $\omega_z = 2\pi \times 20$~kHz. 
	}
	\label{Fig_state_ratio_vs_temp}
	\end{figure}

Note also that as the temperature decreases the fraction of nodal (slow decaying) states decreases (see Fig.~\ref{Fig_state_ratio_vs_temp}). Indeed, in the few or single state limit, there will be almost no nodal states in the ensemble. Consequently, the photoassociation dynamics will be dominated by the ground state dynamics and will follow a fast decay. We ran our numerical simulations in this regime, and this is indeed what they show. This is additional evidence that the lower simulation temperature could contribute to the disagreement between the simulated and experimentally-observed slow decay timescale.

\begin{figure}[t!]
	\begin{center}
	\vspace{1em}
  \includegraphics[width=\figWidth mm]{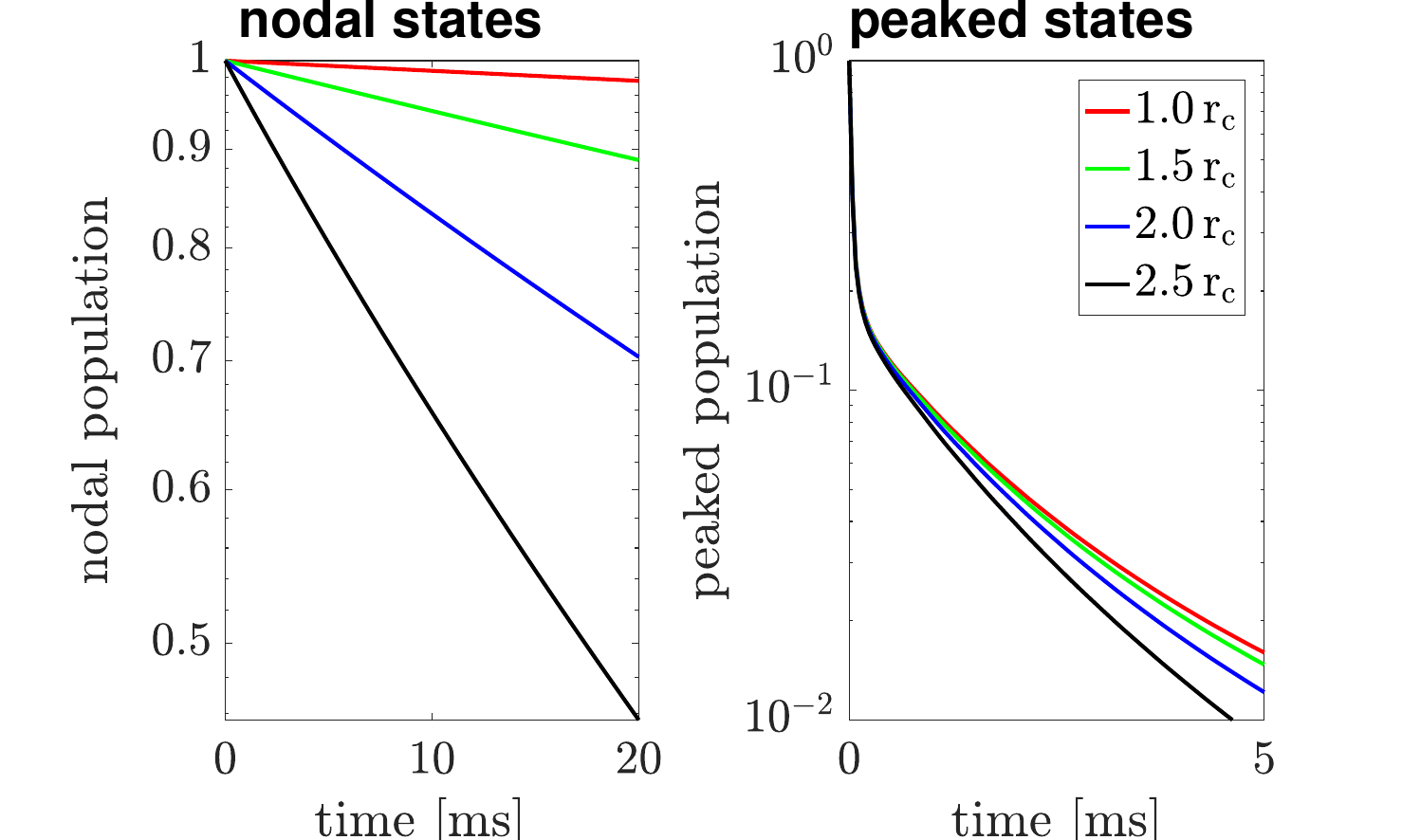}
  \end{center}
	\caption{Change of the photoassociation dynamics at different hard-sphere radii, $R$, in units of the Condon radius $r_c$\,=\,4.3\,nm, separated by state symmetry. For these simulations $\Gamma = K / \mathcal{V}_\text{sphere}$ where $K = 3.59 \times 10^{-16}$~m$^3$/s and $\mathcal{V}_\text{sphere} = \frac{4}{3} \pi R^3$.}\label{fig:sim_n_p_rc}
\end{figure}
\newpage
Figure~\ref{fig:sim_n_p_rc} shows how increasing the range of the absorbing hard-sphere potential changes the simulated decay dynamics of the peaked and nodal states. Note that we have adjusted $\Gamma$ such that the integral over $V_\text{PA}$  remains constant. Although increasing the potential range minimally alters the fast timescale (peaked state) dynamics, it does considerably increase the decay rate for the slow timescale (nodal state) dynamics. This suggests that underestimating the interatomic distance at which loss due to photoassociation occurs will result in a too small slow decay rate. 

\end{document}